\documentclass[12pt,a4paper]{article}
\usepackage{graphics}
\begin{document}

\begin{titlepage}
%
%
\begin{flushright}
    \large
    ICEPP Report-2008/002\\
    9th September 2008
\end{flushright}

\bigskip

\bigskip

\bigskip

\begin{center}
{\LARGE\bf First test of $O(\alpha ^{2})$ correction of the orthopositronium decay rate}
\end{center}

\bigskip

\bigskip

\bigskip

\begin{center}
{\large Y.~Kataoka, S.~Asai  and T.~Kobayashi}
\end{center}

\noindent International Center for Elementary Particle Physics, 
and Department of Physics, Faculty of Science, University of Tokyo,
Hongo, Bunkyo-ku, Tokyo 113-0033

\begin{abstract}

Positronium is an ideal system for the research 
of the bound state QED.
New precise measurement of orthopositronium 
decay rate has been performed with 
an accuracy of 150~ppm.
This result is consistent with the last three results
and also the 2nd order correction.
The result combined with the last three is
7.0401$\pm$0.0007$\mu \mathrm{sec}^{-1}$ (100~ppm), which
is consistent with the 2nd order correction and 
differs from the 1st order calculation by 2.6$\sigma$\@.
It is the first test to validate the 2nd 
order correction.

\end{abstract}

\end{titlepage}

\newpage

\section{Introduction}

Positronium (Ps), the bound state of 
an electron and a positron,
is a purely leptonic system
and the triplet ($1^{3}S_{1}$) state of Ps, orthopositronium(o-Ps),
decays slowly into three photons.
Precise measurement of this decay rate 
gives us direct information about quantum electrodynamics(QED)
in bound state.

Three precise measurements\cite{GAS87,GAS89,CAV90} 
of the o-PS decay rate were performed,
in which reported decay rate values were much larger, i.e., 5.2 -- 9.1 experimental 
standard deviations, than a QED prediction\cite{ADKINS1,log1}
($7.0382~\mu \mathrm{sec}^{-1}$) corrected up to $O(\alpha)$ and $O(\alpha^2 \log(1/\alpha))$\@.
This discrepancy has been referred as `orthopositronium lifetime puzzle', 
and was long-standing problem.
In an effort to clarify the discrepancy, various 
exotic decay modes have been searched for,
without any evidence so far\cite{ex1,ex2,ex3,ex4,ex5}.

It was found~\cite{ASAI95,PEKIN} that thermalization process of the 
produced o-Ps is much slower than its lifetime, 
and the unthermalized o-Ps becomes the serious 
systematic error in all the previous measurements.
With new measurement\cite{ASAI95}, 
we obtained decay rate of $7.0398(29)~\mu \mathrm{sec}^{-1}$, which
was consistent with the first order QED calculation,
and quite differ from the previous results.
In this method, the thermalization process was directly measured
and taken into account. 
This result has been confirmed by two different and more accurate~(200~ppm)
measurements\cite{JIN,MPOL} and the `lifetime puzzle' has been 
solved\cite{ASAI02}\@.
Since the same results are obtained with different two 
methods(\cite{ASAI95,JIN} \cite{MPOL}),
these methods have the potential to measure decay rate more precisely. 

In our previous two measurements\cite{ASAI95,JIN},
the NaI(Tl) and CsI(Tl) scintillators were used, whose
decay constants were long and rise times were also
not fast.
It was the reason that the timing performance 
of the previous measurements were limited.
Furthermore, a low activity source of the positron only can be used
in order to suppress pileup, since they have long decay constants.
Many types of newly developed scintillators have been tested,
and the YAP scintillator is used for this measurement.    
We can obtain good time performance and we can use positron source with an activity of 11 kBq.
In order to reduce the trigger of the useless events,
anti-trigger system is introduced.
If the positron annihilate into 2$\gamma$ on the vacuum vessel, trigger signal is vetoed.

Non-relativistic-QED approximation (NRQED) has been developed recently, 
and it is useful to calculate the higher order correction of the bound state.
This approach can be used for not only QED but also QCD, in which 
the higher order corrections are crucial due to the strong coupling constant.  
The second order correction $O(\alpha^2)$\cite{ADKINS2}, 
whose contribution is about 243~ppm, 
has been performed in 2000 and 2002, 
and the more accurate measurement is necessary 
to examine  this higher order correction.
The logarithmic corrections $O(\alpha^3 \log(1/\alpha))$ and $O(\alpha^3 \log^2(1/\alpha))$
are also calculated and these contributions are found to be small, 3--4~ppm\cite{log2,log3}.
In this paper, we focus on the validation of the $O(\alpha^2)$ correction. 

\section{Experimental setup}

Figure~1 shows a diagram of the experimental setup\cite{kataD};
a ${}^{68}{\rm Ge-Ga}$ positron source with the 
strength of 11 kBq, being sandwiched between two sheets of plastic 
scintillators(NE102 thickness=200$\mu$m) 
and held by a cone made of aluminized mylar. 
The cone was situated at the center of a cylindrical vacuum container
made of $1~mm$-thick plastic scintillators and glass, 
being filled with ${\rm SiO_2}$ aerogel~(RUN-I) or powder~(RUN-II),
and evacuated down to $1\times10^{-2}$ Torr. 
Density of the aerogel and powder is 0.03 $\mathrm{g/cm^{3}}$ both,
and the surface of the primary grain are replaced into hydrophobic
in order to remove the electric dipole of the OH-. 
The sizes of the primary grain are 10 and 16 nm 
for the aerogel and powder, respectively.
Thus  the mean free path of the collision between positronium
and ${\rm SiO_2}$ are different,
and it makes difference in the thermalization process 
and the pick-off probability.

\begin{figure}[!h]
\begin{center}
  \resizebox{0.70\textwidth}{!}{
    \includegraphics{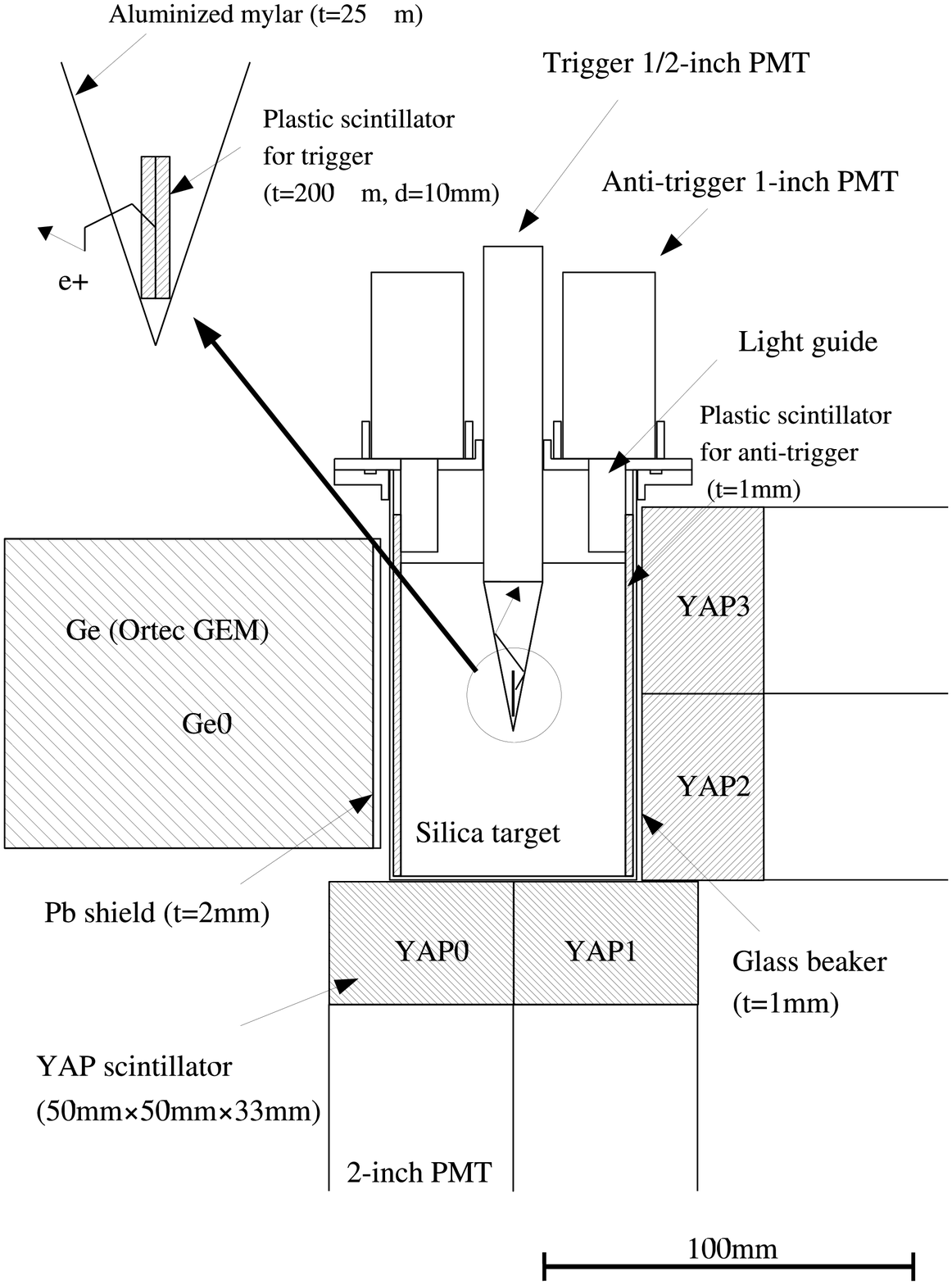}
  }
  \caption{Schematic diagram of apparatus.}
\end{center}
\end{figure}

The emitted positrons pass through thin scintillator,
transmitting a light pulse to 1/2-inch  
photomultiplier(Trigger PMT:Hamamatsu H3165P-10VT) 
and forming Ps when stopped in the silica powder or aerogel.
The signal of the Trigger PMT is used for start signal(t=0) of all TDCs.
Since the kinetic energy of the positron 
is high ($E_{end}$=1.9MeV: ${}^{68}{\rm Ge-Ga}$)
about 40\% of the positron pass through the silica and annihilate at the 
vacuum container.
If the positron deposit energy($>$100~keV) 
in the $1~mm$-thick plastic scintillators, inside wall of 
the vacuum container, the produced trigger signal
is vetoed to remove these annihilate events in the wall.
A light pulse from the $1~mm$-thick plastic scintillators
is guided into two 1-inch photomultipliers(Anti-trigger PMT:Hamamatsu R1924P),
and gives the anti-trigger signal.

Three high-purity coaxial germanium detectors (Ortec GEM38195) 
precisely measured the thermalization process.
One output from the preamplifier of each germanium detector
is fed through a timing-filter amplifier (Ortec 474), whose integration 
and differential times are set to 20 and 500 nsec, respectively.
Outputs of the timing-filter amplifier are put into three 
leading-edge discriminators, in which different threshold 
values are set at 30, 50 and 150~keV, and 
these three timings are recorded individually by TDCs.
The signal shape of the coaxial germanium detectors depends
on the position at which energy is deposited, since the electric field 
for charge collection is complicated.
Time walk correction can be applied 
and correct timing is obtained using these three timings.
The good time resolution of (4~nsec) is obtained for $E_{\gamma}~$480--511~keV.

Another output from  the preamplifier of each germanium detector
is fed into a spectroscopy amplifier(Ortec 673), and 
the signal is integrated with the time constant of 6$\mu \mathrm{sec}$.
The output is recorded with peak-hold ADC(Hoshin C011).
Energy resolutions were measured using several line $\gamma$ sources, 
with typical resultant values of 0.5~keV at 514~keV. 
A $2~mm$-thick lead sheet is placed in front of each germanium detector 
to eliminate the contribution of the 
two low-energy $ \gamma $'s from the three-photon decay of o-Ps.

Four YAP($\rm {YAlO_{3}}$: Ce doped: size is  50$\times$50$\times$33~mm) 
scintillators simultaneously measured the 
time and energy information from each decay.
2-inch photomultiplier(Hamamatsu R329-02) is connected to each crystal.
The charge ADCs are used to measure the energy information. 
A single leading-edge discriminator (threshold 70~keV) 
is used for the timing, 
and the time walk correction is applied as a function 
of the energy information. 
Since the decay constant of the YAP is fast(30~nsec) and the slow component
of the scintillation is negligibly small,
the event pileup probability is  small and the time walk correction
can be performed perfectly as a function of the deposited energy. 
The good time resolution of 1.2~nsec is obtained for $E_{\gamma}>$150~keV.
YAP is an ideal scintillator for the timing measurements; 
high density (5.27$\mathrm{g/cm^3}$), high light yield (40\% of NaI(Tl)), 
low contamination of the radio activities, 
the fast rise time($<$10 nsec), 
and the fast decay constant(30 nsec).
Detail studies of the various scintillators are summarized in \cite{kataD}.

A direct clock (2 GHz) count type TDC is used for both
the germanium detectors and the 
YAP scintillators. 
The time resolution and the full range are 0.5~nsec and 32~$\mu \mathrm{sec}$(16bit), respectively.
The used clock is calibrated with an accuracy within 1~ppm
and stable ($<$ 2.5~ppm within $-10$ -- 50$^{o}C$).
The integrated non-linearity is negligibly small and 
the differential non-linearity is 0.1\% for 4bin.

Data are taken with the CAMAC system for 2.3 months(RUN-I) and 
3.1 months(RUN-II),
and the trigger rates before the anti-trigger 
are 10.8 and 6.8 kHz, respectively. 
Total observed events of $\beta^{+}$ emission 
are $1.4 \times 10^{10}$ and $1.6 \times 10^{10}$.
The energy calibration and time calibration
has been performed for every 6 hours.
Area temperature was maintained within $\pm 0.5^{o}C$ to
ensure the stability of energy and time calibration. 

As some fraction of o-Ps inevitably results in `pick-off' annihilation 
due to collisions with atomic electrons of the target material, 
the observed o-Ps decay rate 
$\lambda_{obs}$ is a sum of the intrinsic o-Ps 
decay rate $\lambda_{3\gamma}$ and the pick-off 
annihilation rate into $2\gamma$'s, 
$\lambda_{pick}$, i.e.,

\begin{equation}
\lambda_{obs}(t)=\lambda_{3\gamma}+\lambda_{pick}(t).
\end{equation} 

Since the velocity of the Ps slows down slowly due to the elastic collision
to the material, pick-off probability depends on the time after produced
as shown in Figures.2.
The energy distribution of photons emitted from the 3-body decay is 
continuous below the steep edge at 511~keV, 
whereas the pick-off annihilation is 2-body which produces 
a 511~keV monochromatic peak. 
Energy and timing information are simultaneously
measured with high-energy resolution germanium detectors such 
that $\lambda_{pick}(t)/\lambda_{3\gamma}$ can be determined
from the energy spectrum of the emitted photon. 
Once a precise thermalization function is obtained, $\lambda_{pick}(t)$ will 
contain all information about the process. 
The population of o-Ps at time $t$, 
$N(t)$ can be expressed as \begin{equation}
N(t)=N_0' \exp\left(-
\lambda_{3\gamma}\int^{t}_0\left(1+\frac{\lambda_{pick}(t')}
{\lambda_{3\gamma}}\right)dt'\right).
\end{equation}
Providing the ratio is determined as a function of time,
the intrinsic decay rate of o-Ps, $\lambda_{3\gamma}$, can be directly obtained
by fitting the observed time spectrum.

\section{Analysis}

The ratio $\lambda_{pick}(t)/\lambda_{3\gamma}$ is 
determined using the energy spectrum measured by 
the germanium detectors. The energy spectrum of the o-Ps decay sample, 
referred to as the {\it o-Ps spectrum}, is 
obtained by subtracting accidental contributions 
from the measured spectrum. 
The $3\gamma$-decay continuum spectrum is estimated using Monte Carlo
simulation in which the geometry and various material distributions
are reproduced in detail\cite{kataD}.
For every simulated event, three photons are generated.
Successive photoelectric, Compton, or Rayleigh scattering interactions 
of every photon are then followed through the materials 
until all photon energy is either deposited or escapes  from the detectors. 
The response function of the detectors is determined based on the measured spectrum of 
monochromatic $\gamma$-rays emitted from ${}^{152}{\rm Eu}$, ${}^{85}{\rm Sr}$, and 
${}^{137}{\rm Cs}$.
These material and detector effects are taken into account precisely, and the 
{\it $3\gamma$-spectrum} is obtained. These are continuous distributions, 
and  is normalized to the observed o-Ps spectrum with 
the ratio of event numbers  within the region (480--505~keV). 

Figures 2(a)(b) show the o-Ps spectra and 3-$\gamma$ spectra observed in
RUN-I(Aerogel) and II(powder), respectively.
The pick-off 2-$\gamma$ spectra, which are obtained from the o-Ps spectra after 
subtracting the $3\gamma$-spectra, are also superimposed in the same figures. 
The $\lambda_{pick}(t)/\lambda_{3\gamma}$ can be calculated directly from the 
ratio of the pick-off 2-$\gamma$  and 3-$\gamma$ spectrum, 
the ratio does not depend on the absolute values of the detection efficiencies.
The relative value of the detection efficiency $\epsilon_{511~keV}/\epsilon_{480-505~keV}$
is estimated with the monochromatic $\gamma$-rays emitted from
${}^{152}{\rm Eu}$, ${}^{85}{\rm Sr}$, and 
${}^{137}{\rm Cs}$.  
The calculations of the $\lambda_{pick}(t)/\lambda_{3\gamma}$ are
performed for the various time windows,
and its time dependence are shown in Figure 2(c)(RUN-I) and (d)(RUN-II).
Horizontal axis of the figures is the time between the positrons emission and 
decay, this slope shows the thermalization process of the positronium.
It takes much time than the lifetime to well thermalized, as we have already shown
in the previous measurements\cite{ASAI95,JIN}.  
Since the collision rate in RUN-I is higher than that in RUN-II, 
slopes of the thermalization process are different as shown in these figures.

\begin{figure}[!h]
  \resizebox{1.0\textwidth}{!}{
    \includegraphics{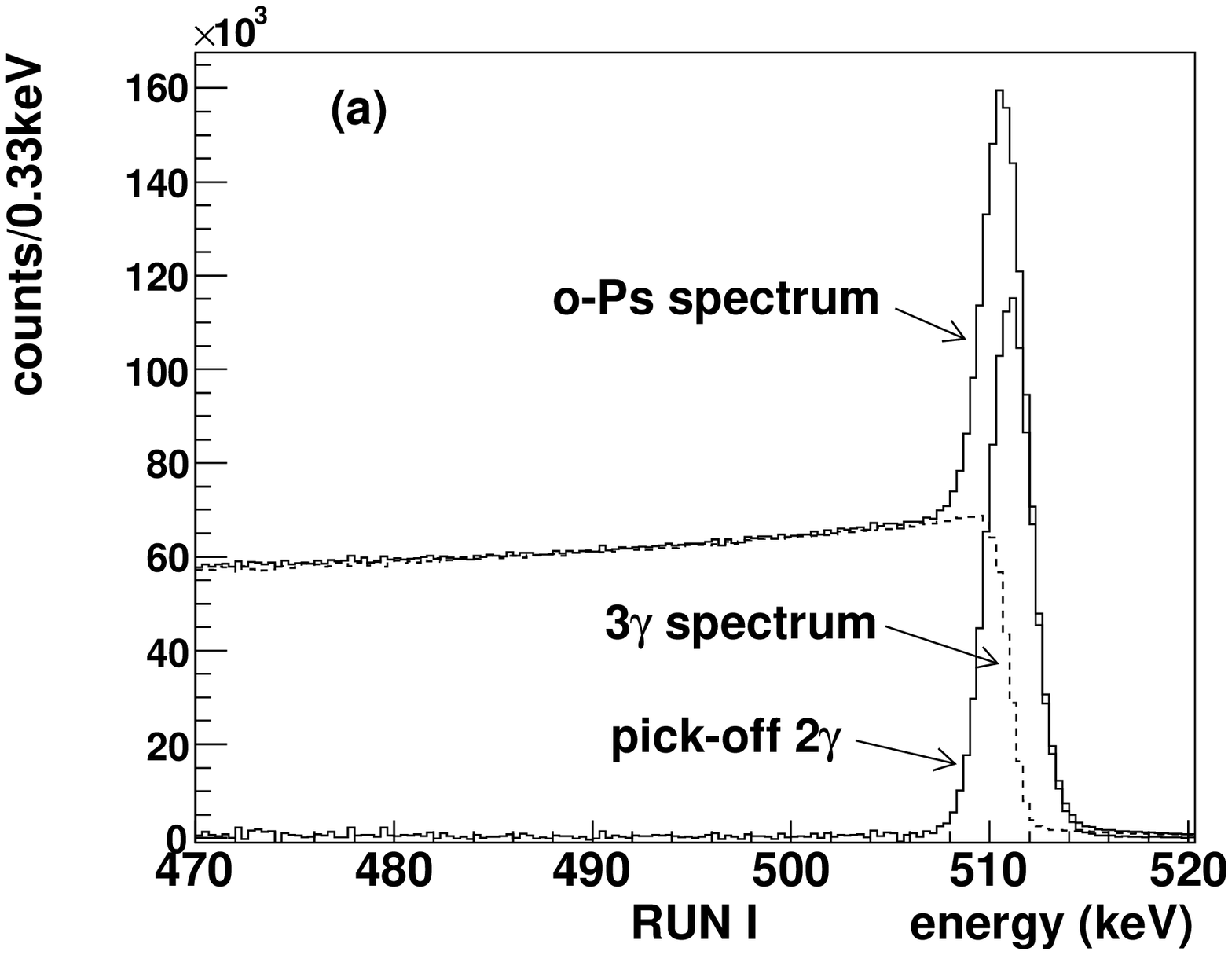}
    \includegraphics{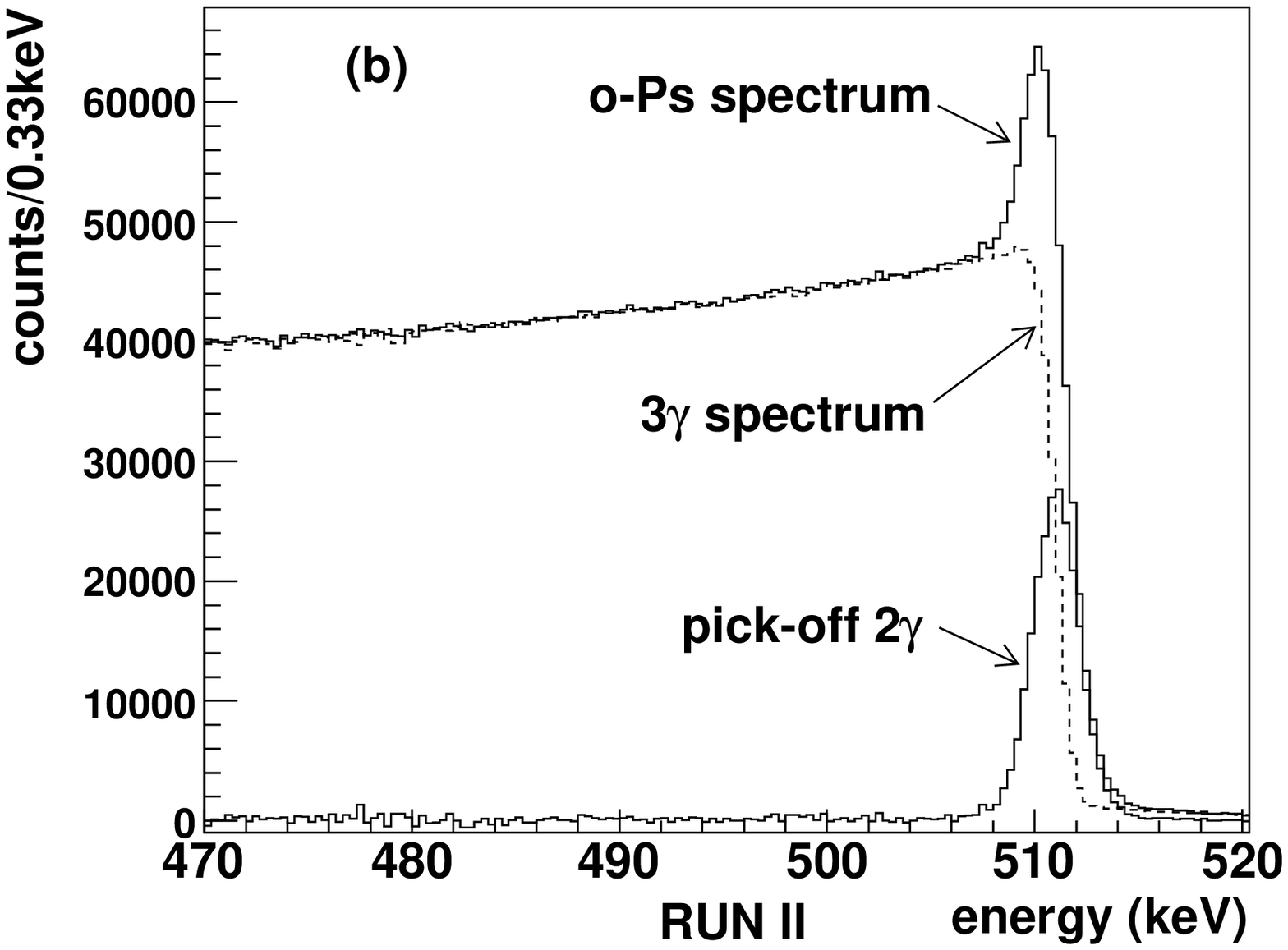}
  }
  \resizebox{1.0\textwidth}{!}{
    \includegraphics{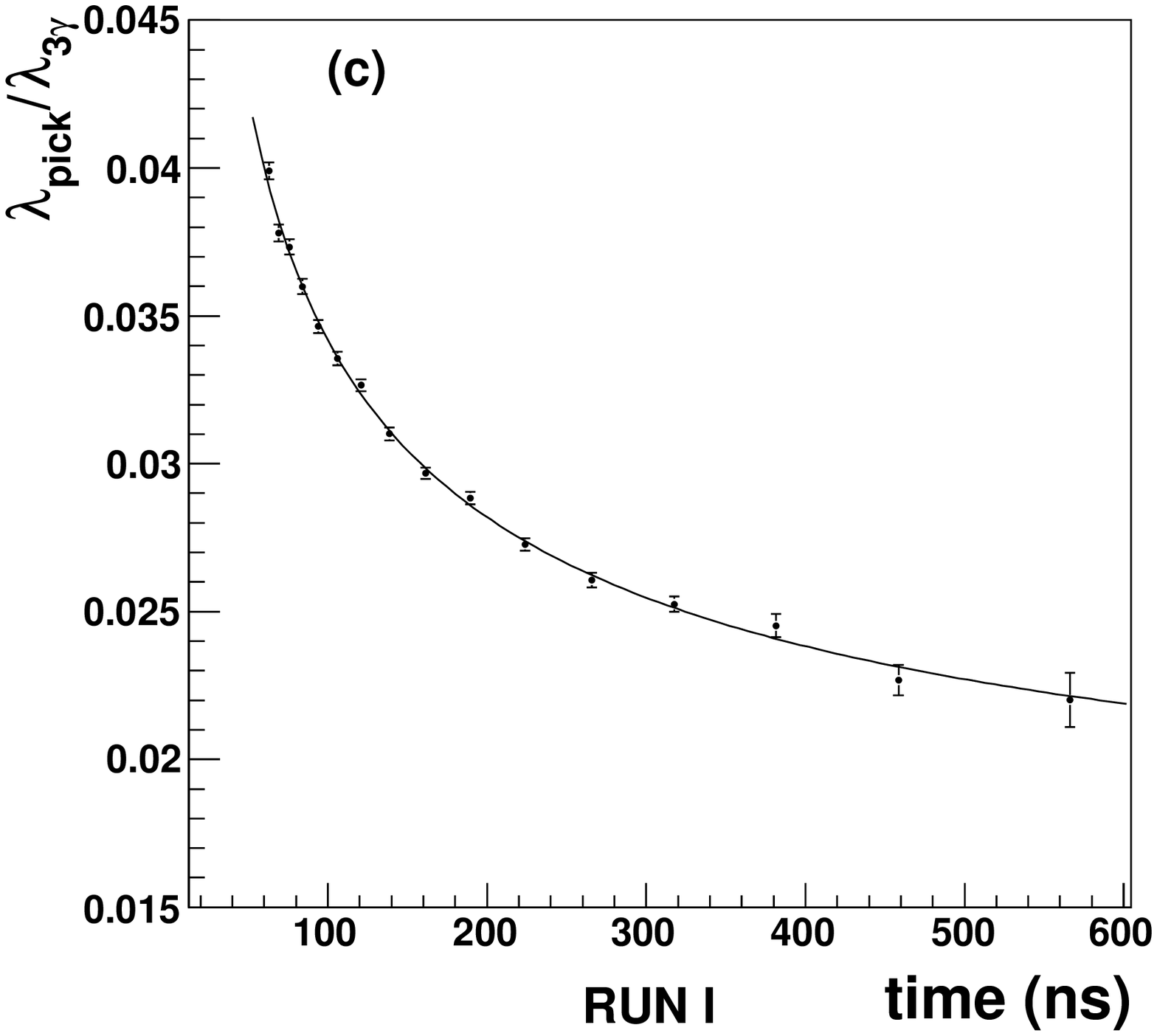}
    \includegraphics{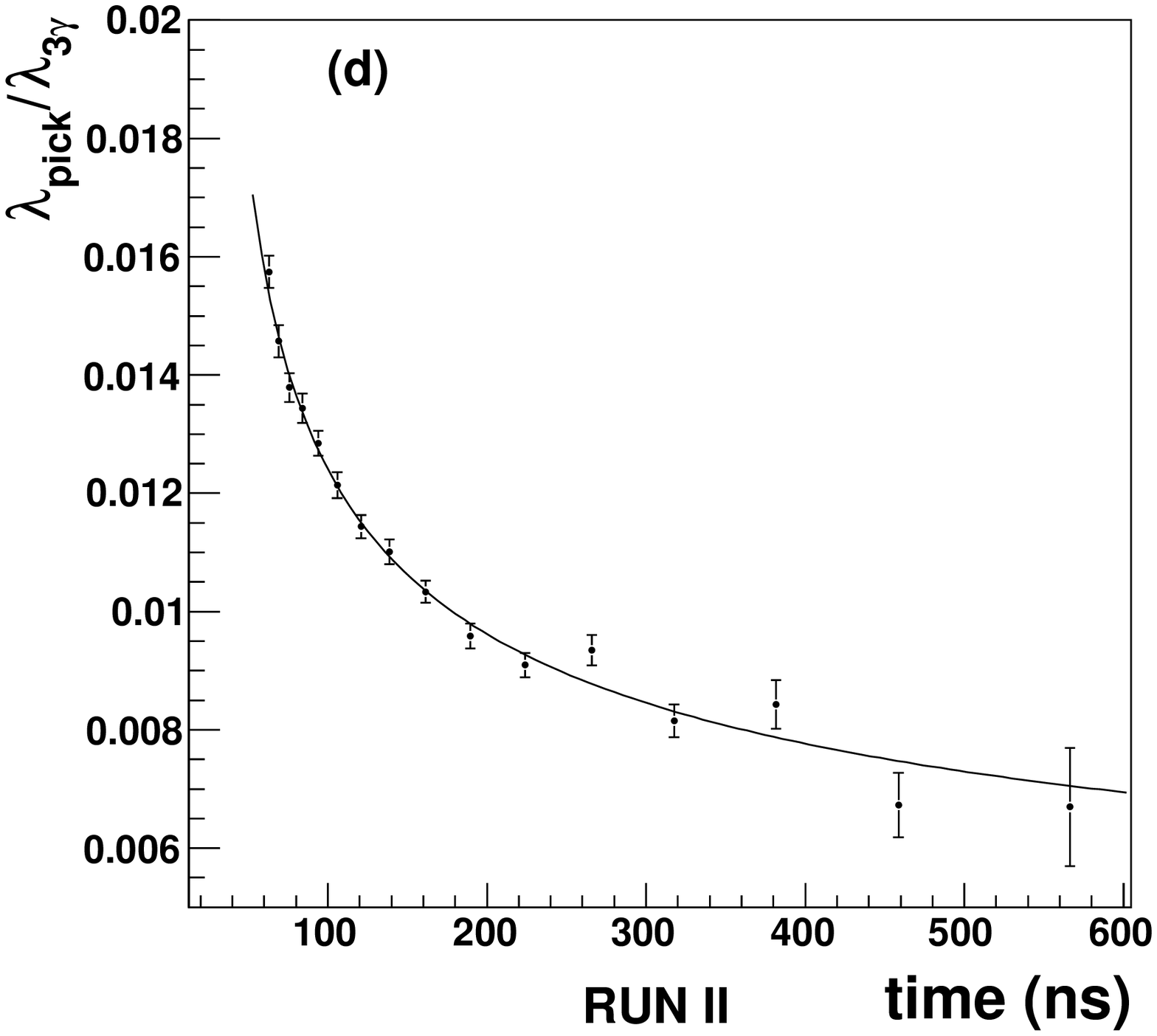}
  }
  \caption{(a) Energy spectra of o-Ps decay $\gamma$'s obtained by germanium detectors in 
RUN-I(Aerogel). 
The solid line represents data points in a time window of 60--700~nsec, and the dotted 
line shows the $3\gamma$-decay spectrum calculated by the Monte Carlo 
simulation. Pick-off spectrum obtained after subtracting the $3\gamma$ contribution 
from the o-Ps spectrum is superimposed. 
(c) The ratio $\lambda_{pick}(t)/\lambda_{3\gamma}$ are plotted as a function of 
time (RUN-I). Only the statistical errors are shown and the solid lines 
represent best fit results obtained.
(b),(d) the same figures for RUN-II(Powder) }
\end{figure}

Figures~3(a) and (b) show the time spectra measured 
with the YAP scintillators in RUN-I and -II. 
A sharp peak of the prompt annihilation is followed by the exponential decay 
curve of o-Ps and then the constant accidental spectrum. 
Time resolution of 1.2 nsec is obtained and 
the o-Ps curve is widely observed over $1.0~\mu \mathrm{sec}$.
We fit resultant time spectrum using the least square method
with the following function;
\begin{equation}
N_{obs}(t)=\exp(-R_{stop}t)\left[\left(1+\frac{\epsilon_{pick}}
{\epsilon_{3\gamma}}\frac{\lambda_{pick}(t)}{\lambda_{3\gamma}}\right)N(t)
+C\right], 
\label{eq:tspec_obs}
\end{equation} 
where $\epsilon_{pick}$ and $\epsilon_{3\gamma}$ are 
the detection efficiencies for pick-off 
annihilation and $3\gamma$ decays, and $R_{stop}$ 
is an experimental random counting rate 
representing the fact that time interval measurement 
always accept the first $\gamma$ as a stop signal. 
$\lambda_{pick}(t)/\lambda_{3\gamma}$ is about $1\%$ 
due to the low-density of the ${\rm SiO_2}$ powder or aerogel.
This means that
the ratio of error propagation to the decay rate is suppressed by a factor of 100. 
It is a merit of our method, and an accuracy of 100~ppm can be achieved
if $\lambda_{pick}(t)/\lambda_{3\gamma}$ is measured with an accuracy of 1\%,
which is the realistic.

\begin{figure}[!h]
  \resizebox{1.0\textwidth}{!}{
    \includegraphics{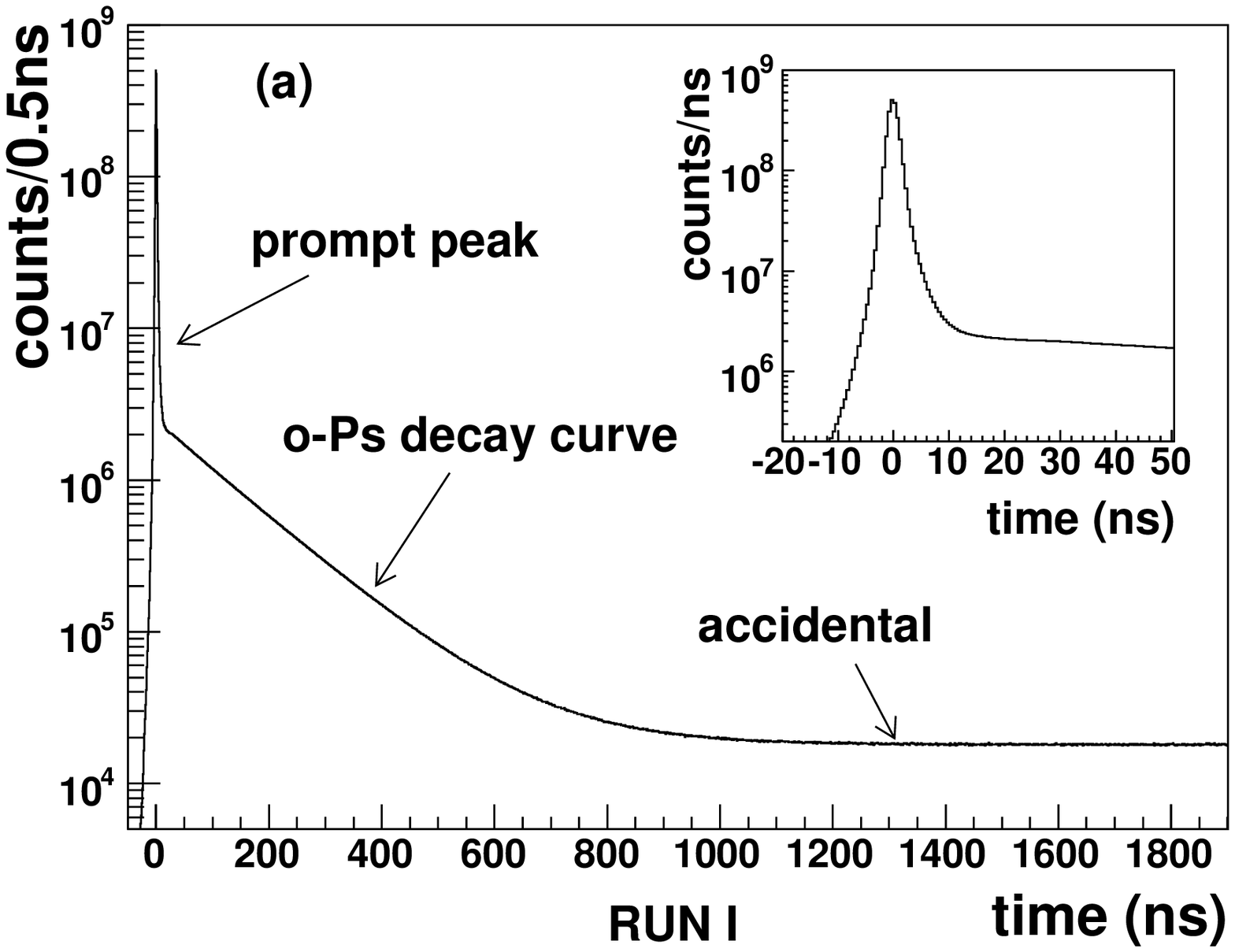}
    \includegraphics{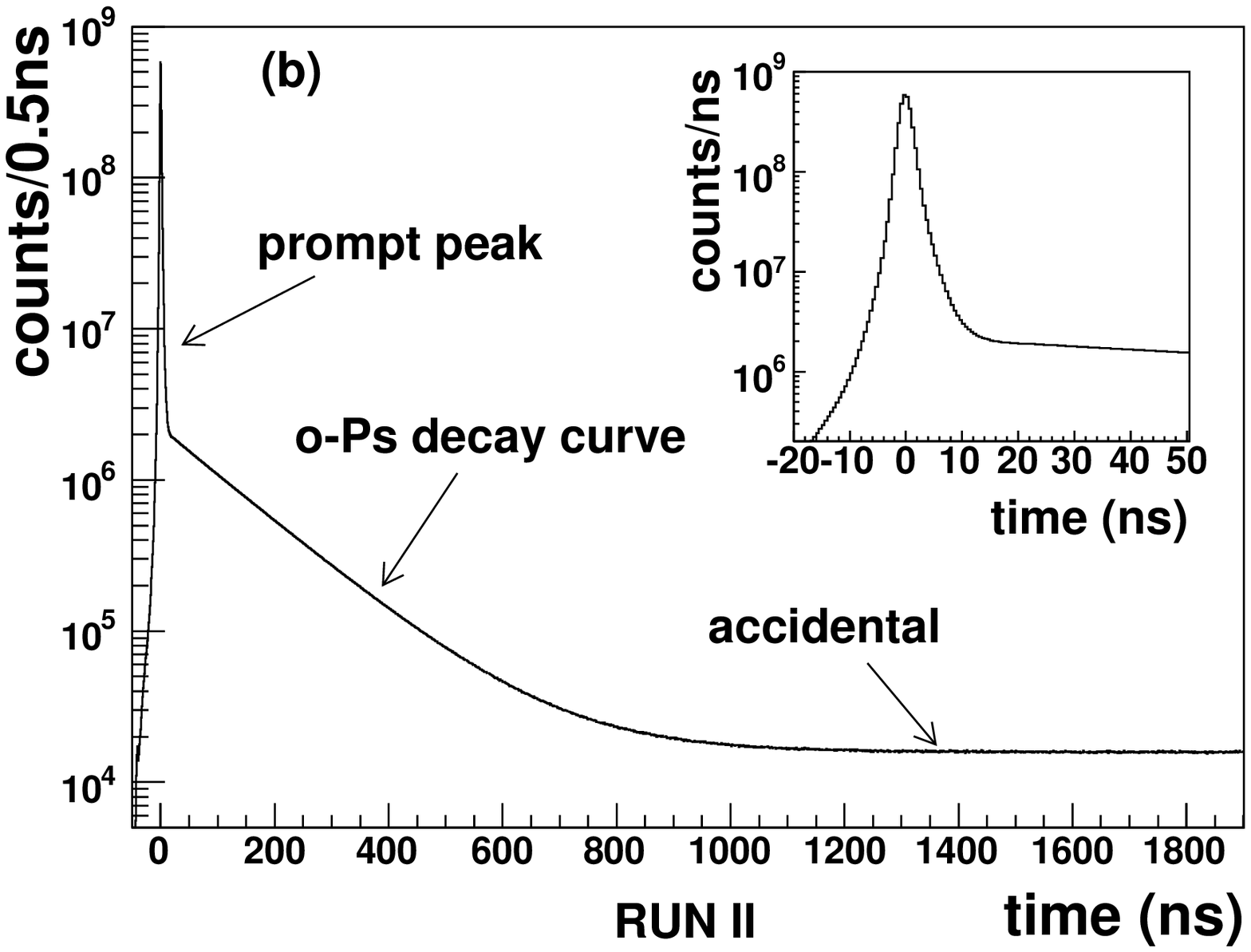}
  }
  \resizebox{1.0\textwidth}{!}{
    \includegraphics{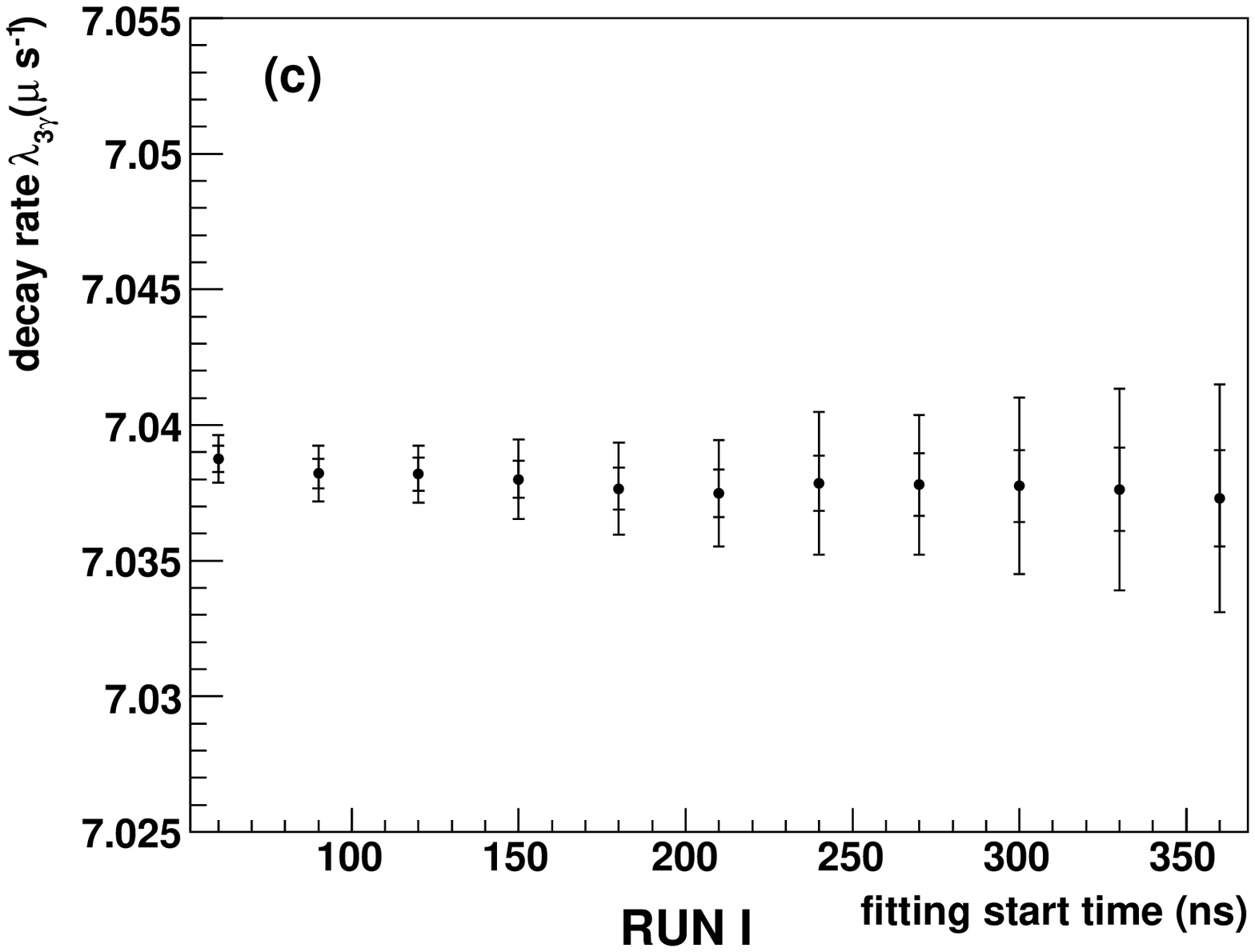}
    \includegraphics{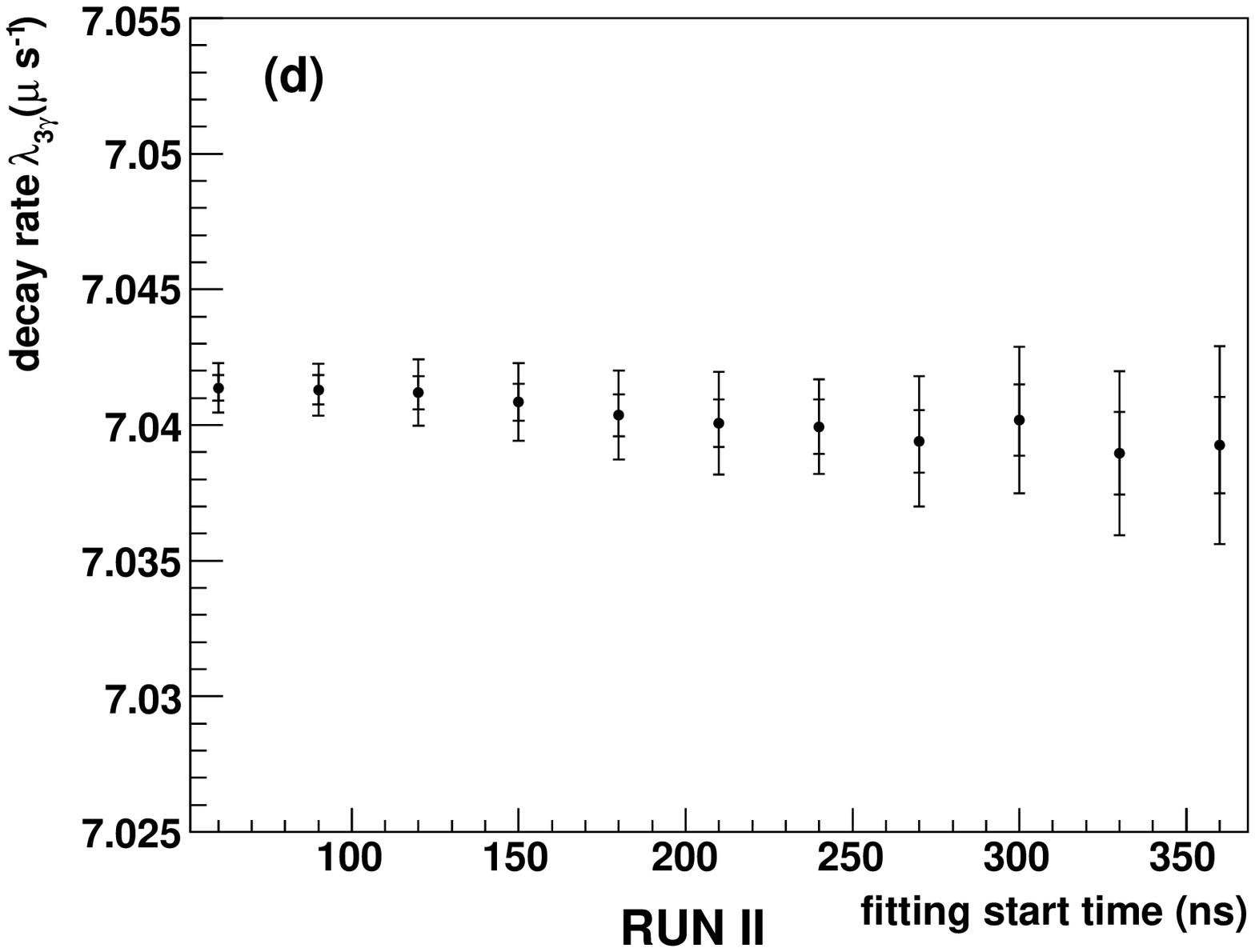}
  }
  \caption{(a) Timing spectrum measured by the YAP scintillators 
within an energy window above 150~keV (RUN-I).
The figures in the right corner show
the magnified view of the prompt peak. 
(b) The same timing spectrum in RUN-II.
(c)(d) The fitted results are plotted as a function of the fitting start time 
for RUN-I and -II.}
\end{figure}

The fitting procedure has been performed for the various fitting region.
Figures~3(c) and (d) show the fitted results as a function of the fitting 
start time, end point is fixed at t=3600 nsec.
If the ratio $\lambda_{pick}(t)/\lambda_{3\gamma}$ would be mismeasured,
the fitted results becomes unstable against the fitting region.
The fitted results are stable against the fitting regions as shown in these 
figures and the values with the fitting start time of 60 nsec are used for both run.
The obtained decay rates\cite{kataD} are 
$\lambda_{3\gamma}=7.03876\pm0.0009(stat.)~\mu \mathrm{sec}^{-1}$ for RUN-I and 
$7.04136\pm0.0009(stat.)~\mu \mathrm{sec}^{-1}$ for RUN-II, 
which are consistent with each other. 

Estimates of various systematic errors are summarized in Table~1.

\begin{center}
\begin{table}
\begin{tabular}{|l|r|r|}
\hline
 Source & RUN-I(ppm) & RUN-II(ppm) \\ \hline \hline
TDC non linearity & $<\pm$ 15 & $<\pm$ 15 \\ \hline
Pile up           & $< +$ 10 & $< +$ 10 \\ \hline
Pickoff estimation &    &   \\ 
 (1) 3$\gamma$ subtraction & $<\pm$ 89 & $<\pm$ 91 \\
 (2) Ge efficiency         & $<\pm$ 33 & $<\pm$ 28 \\
 (3) YAP efficiency        & $<\pm$ 64 & $<\pm$ 19 \\ \hline
Physics source &    &   \\ 
 (1) Zeeman effect         & $< -$ 5 & $< -$ 5 \\
 (2) Stark effect          & $< +$ 3 & $< +$ 4 \\
 (3) 3$\gamma$ annihilation & $< -$ 91 &$< -$ 33 \\ \hline \hline
Total     & -147 and + 115 & -104 and +98 \\ \hline 
\end{tabular}
\caption{Summary of the systematic errors}
\end{table}
\end{center}

The predominant contribution to total systematic error is produced 
by uncertain normalization of the 3-$\gamma$ spectrum.
That is, the number of pick-off events are determined by subtracting 
the normalized 3-$\gamma$ spectrum of Monte 
Carlo simulation from the o-Ps spectrum, 
where changing the normalization factor affects the 
$\lambda_{pick}(t)/\lambda_{3\gamma}$ values and 
eventually propagates to the final result. 
Since the sharp fall-off of the 3$\gamma$-spectrum at 511~keV is solely 
produced by the good germanium energy resolution of $\sigma= 0.5$~keV, 
this subtraction only affects the lower side 
of the pick-off spectrum such that 
improper subtraction results in asymmetry of the pick-off spectrum shape. 
Comparison of the asymmetries of the pick-off peak shape 
and the prompt peak annihilation spectrum is a good parameter for estimating 
this systematic error. 
The errors are about 90~ppm for both measurements.

The relative value of the detection 
efficiency $\epsilon_{511~keV}/\epsilon_{480-505~keV}$
is used to estimate $\lambda_{pick}(t)/\lambda_{3\gamma}$.
The relative efficiency is directly measured using the $\gamma$-rays 
and there are two other uncertainties in the relative efficiency. 
(1) If the density of silica power(RUN-II) changes by 10\% 
during the run time of 3 months,
the distribution of the decay points of orthopositronium changes
and detection efficiency change by 0.12\%.
(2) Some part of the germanium signal has very 
slow rise time(SRT), and these events are rejected 
using the timing information of three different thresholds. 
This SRT rejection efficiency are estimated using the real data.
The uncertainties of the estimation are 0.08\%(RUN-I) and 0.09\%(RUN-II).
These two uncertainties are propagated to 
the fitted valeus of $\lambda_{3\gamma}$, 
about 30~ppm for both runs.

The fitting function includes the relative efficiency of the YAP scintillator
as shown in the equation(3). 
This relative efficiencies are also estimated with real data using 
the $\gamma$-rays emitted from $^{85}\mathrm{Sr}$(514~keV). 
An uncertainties of the relative efficiency are 2\%(RUN-I) and 1.3\%(RUN-II),
and they contribute to the errors on the fitted  valeus of $\lambda_{3\gamma}$
by 64~ppm(RUN-I) and 19~ppm(RUN-II). 

The Stark shift stretches the lifetime of Ps atoms. 
A calculation shows that the shift is 
proportional to a square of the effective electric field $E$ such that 
$\triangle\lambda_{3\gamma}/\lambda_{3\gamma}=248\cdot(E/E_0)^2$, 
where $E_0 = m_e^2e^5/\hbar^4 \approx 5.14\times10^9~\mathrm{V/cm}$. 
Silanol functional groups on the surface of the grain 
behave as an electrical dipole moment creating an effective field around the grains. 
Average densities of Silanol are measured to be  $0.44 \mathrm{nm}^{-2}$ and
the effective field can be analytically calculated such that the contribution to the o-Ps decay 
rate is determined to be -5~ppm.
These estimations were confirmed  using results from precise hyper-fine-structure (HFS)
interval measurements of ground state Ps in silica powder\cite{HFS}, 
where the interval is proportional to the size of Stark effect. 
Considering the difference in powder densities used, 
the HFS results are consistent with our estimation.

Error contribution due to the Zeeman effect is estimated
using the measured absolute magnetic field around the positronium assembly (-5~ppm). 
Since the 3-$\gamma$ pick-off process can only occur at a certain 
ratio, and the calculated relative frequency.
On the other hand, the pick-off events by the {\it spin-flip} effect 
do not have the 3$\gamma$ decay process.
If the contribution of the {\it spin-flip} is neglected, 
the 3$\gamma$ pick-off events increase by 0.3~\% and,
being -91~ppm for RUN-I and -33~ppm for RUN-II. 

The above discussed systematic errors are regarded 
as independent contributions such that the total 
systematic error can be calculated as their quadratic sum, 
resulting in -147~ppm, +115~ppm for RUN-I 
and -104~ppm, +98~ppm for RUN-II.
The combined result with systematic error is 
$\lambda_{3\gamma}=7.0401\pm0.0006(stat.)^{+0.0007}_{-0.0009} (sys.) ~\mu \mathrm{sec}^{-1}$,
and total error is 150~ppm.
Correlation of the systematic errors are taken into account.

\section{Results and discussion}

Figure 4 shows the summary of the measured decay rate after 1995,
three results are obtained with our method and 
one result is obtained using thin polymer, with which the produced
positronium is very slow and the effect of the thermalized 
positronium is suppressed.
These four results are consistent with each other.
And the combined value 
is $\lambda_{3\gamma}=7.0401\pm0.0007~\mu \mathrm{sec}^{-1}$
and shown in the red arrow in the figure.
The correlations of the systematic errors 
are carefully taken into account.
The combined result is consistent with the all four results 
and accuracy is 100~ppm.
This result is consistent with the 
O($\alpha ^{2}$) correction\cite{ADKINS2},  
and differs from the only up to O($\alpha$)  by 2.6$\sigma$. 

\begin{figure}[!h]
\begin{center}
  \resizebox{0.7\textwidth}{!}{
    \includegraphics{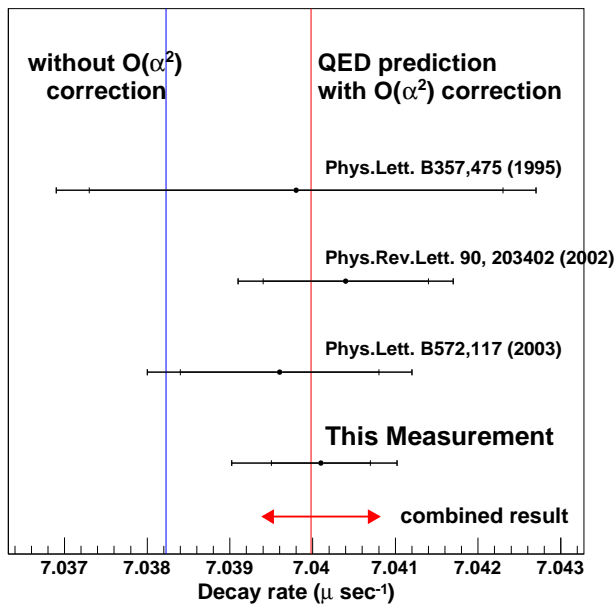}
  }
  \caption{Decay rates measured in the last four experiments
 are listed chronically.
The last arrow shows the combined result of the last four measurements. 
The red and blue lines shows the QED prediction calculated up to O($\alpha^2$) and O($\alpha$), respectively }
\end{center}
\end{figure}

\section{Conclusion}

The decay rate of o-Ps was measured using a 
direct $2\gamma$ correction method in which the thermalization 
effect of o-Ps is accounted for and integrated into the time spectrum 
fitting procedure. 
The new result is 
$\lambda_{3\gamma}=7.0401\pm
0.0006(stat.)^{+0.0007}_{-0.0009} (sys.) ~\mu \mathrm{sec}^{-1}$,
and total error is 150~ppm\cite{kataD}.
It is the most accurate result and agrees well 
with the last three results\cite{ASAI95,JIN,MPOL}. 
The combined result of these four measurements
is 
7.0401$\pm$0.0007$\mu \mathrm{sec}^{-1}$,
which is consistent well with the NRQED prediction 
corrected up to $O(\alpha^2)$ term\cite{ADKINS2},
and differs from the result up to $O(\alpha)$ by 
2.6$\sigma$\@.
It is the first result to validate the $O(\alpha^{2})$
correction.

We thank Prof. M.Kobayashi(KEK) for very useful discussion 
about the scintillators.
Sincere gratitude is extend to Dr.M.Ikeno(KEK) and
Dr.T.Namba(U.Tokyo) for a development of 2 GHz TDC and  
the great support to this measurement.
We are pleased to acknowledge the 
Japan Society for the Promotion of Science
for the financial support.

%

\end{document}